\documentclass[graybox]{svmult}
\usepackage{mathptmx}       
\usepackage{helvet}         
\usepackage{courier}        
\usepackage{makeidx}         
\usepackage{graphicx}        
\usepackage{multicol}        
\usepackage[bottom]{footmisc}
\usepackage{mathtools,amssymb,amsfonts}
\usepackage{xcolor}

\usepackage{comment}

\makeindex             

\newcommand\G{\mathrm{G}}
\newcommand\HH{\mathrm{H}}

\newcommand\LL{\textsc{l}}
\newcommand\RR{\textsc{r}}

\newcommand\F{\mathrm{F}}
\newcommand\EE{\mathrm{E}}
\newcommand\N{\mathrm{N}}

\newcommand\PP{\mathrm{P}}

\DeclareMathOperator*{\ordprod}{\prod\limits^{\vbox to -.5ex{\kern-0.5ex\hbox{$\leftharpoonup$}\vss}}}
\DeclareMathOperator*{\ordprodopp}{\prod\limits^{\vbox to -.5ex{\kern-0.5ex\hbox{$\rightharpoonup$}\vss}}}

\DeclarePairedDelimiter{\ket}{\lvert}{\rangle}
\DeclarePairedDelimiterX{\ketbra}[2]{\lvert}{\rvert}{#1\rangle \langle#2}
\DeclarePairedDelimiterX{\braket}[2]{\langle}{\rangle}{#1\vert#2}

\begin{document}
\title*{The \textit{q}-deformed Haldane--Shastry chain at \textit{q} = \textit{i} with even length}
	\author{Adel Ben Moussa, Jules Lamers and Didina Serban}
	\institute{ABM \at Universit\'e Paris--Saclay, CNRS, CEA, Institut de Physique Th\'eorique, 91191 Gif-sur-Yvette, France, \email{adel.benmoussa@ipht.fr}
    \and DS \at Universit\'e Paris--Saclay, CNRS, CEA, Institut de Physique Th\'eorique, 91191 Gif-sur-Yvette, France, \email{didina.serban@ipht.fr}
 	\and JL \at Deutsches Elektronen-Synchrotron DESY, Notkestra{\ss}e 85, 22607 Hamburg, Germany, \email{jules.lamers@desy.de}
    }

\maketitle

\abstract{In this note we announce some results extending our recent work with A.~Toufik on the free-fermion point $q=i$ of the Haldane--Shastry chain to the case with an even number $N$ of sites. The resulting long-range version of the Heisenberg XX chain may be viewed as a model of fermions with extended $\mathfrak{gl}(1|1)$ symmetry. Unlike for odd $N$, the conserved charges are nilpotent and exhibit Jordan blocks.}

\section{Introduction}
\label{sec:1}

The advent of logarithmic conformal field theories at the end of the last century, with applications in condensed matter theory and high-energy physics, has led to a growing interest in models whose algebraic structures are not completely reducible with respect to their algebras of symmetries \cite{gurarie2013logarithmic, Read_07, rozansky1992quantum}. More concretely, this property is related to conserved charges that are not diagonalisable, exhibiting `non-trivial Jordan blocks'. An example of such a structure arises in the quantum-group invariant open \textsc{xx} chain with an even number of sites. Its symmetries make it a non-semisimple representation for a specialisation of $U_q(\mathfrak{sl}_2)$ at $q=i$ and, in the continuum limit, yields a well-known logarithmic CFT called the free symplectic fermions theory \cite{Read_07}. 

In this note, we announce new results for the free-fermion point of the Haldane--Shastry chain, extending the recent work \cite{bm2024} 
to the case with an even number of sites.
The Haldane--Shastry model \cite{Hal_88, Hal_94} is an exactly solved spin chain with long-range interactions and remarkable properties including explicit eigenvectors and extended (Yangian) spin symmetry \cite{Bernard:1993va, HH+_92}.
It admits an integrable $q$-deformation, similar to how the Heisenberg \textsc{xxz} chain generalises the isotropic Heisenberg \textsc{xxx} chain, called the $q$-deformed Haldane--Shastry ($q$HS) chain \cite{Lamers:2018ypi, Lamers:2020ato, Uglov:1995di}. Amongst others it has a hierarchy of commuting charges, and $q$-deformed (quantum-affine) extended spin symmetry; see the Supplemental Material of \cite{bm2024} for a summary.
While the properties of the $q$HS chain mimic that of HS at generic values of $q$, new phenomena emerge when $q$ is a root of unity.
We consider the case of $q=i$. Just like the Heisenberg \textsc{xxz} chain reduces to the nearest-neighbour \textsc{xx} model at this point, the $q=i$ HS ($i$HS) chain is a model of free fermions, now with long-range hopping. Its properties depend critically on the parity of the number $N$ of sites.

In \cite{bm2024}, we studied the $i$HS chain with odd $N$. Following \cite{Gainutdinov:2011ab}, the spin-chain representation of the Temperley--Lieb algebra can at $q=i$ be related to quadratic combinations of non-unitary versions of Jordan--Wigner fermions. 
We exhibited the first few commuting charges inherited from $q$HS explicitly in the fermionic language. This includes a chiral (parity-odd) hamiltonian that is quadratic in fermions, and a parity-even hamiltonian with quartic interactions implementing a statistical repulsion on the lattice. Both hamiltonians have a linear dispersion relation. The model is not translationally invariant in the traditional sense, yet both hamiltonians commute with a suitably modified `quasi-translation' operator playing a pivotal role in \cite{bm2024}. Although the model is non-unitary, by combined parity and time-reversal symmetry, it has a real spectrum. 

In this note we focus on an even number $N$ of sites. 
The main difference with odd $N$ is that the commuting charges of the $q$HS chain now exhibit singularities at $q=i$ that require regularisation. The renormalised conserved charges all become nilpotent operators. In Section~\ref{sec:2} we give the explicit form of the first few of them in fermionic language and discuss some symmetries. In Section~\ref{sec:3} we discuss examples for low $N$, illustrating that these operators have a non-trivial Jordan-block structure reminiscent of the $U_q(\mathfrak{sl}_2)$-invariant \textsc{xx} chain at $q=i$. In the present case, the Jordan blocks are larger, and we attribute this phenomenon to the extended $U_q(\widehat{\mathfrak{sl}}_2)$ spin symmetry of the $q$HS chain. 
We will elaborate on the extended symmetry its representation in a forthcoming work.

\section{The conserved quantities}
\label{sec:2}

The conserved quantities of the parent model are rational in $q$. In the case of even length $N=2L$, their expressions are singular at $q\to i$, but well-defined conserved charges can be extracted by carefully renormalising the original quantities. In this section, we define these quantities as elements of the Temperley--Lieb algebra, which can be represented as fermionic operators on the Fock space with $N$ fermions, and consider their algebraic properties.

\runinhead{Algebraic setting} 
We define the \emph{free-fermion Temperley--Lieb algebra} by generators $e_1, \dots e_{n-1}$ and relations
\begin{equation} \label{TLrel_freefermion}
	\begin{aligned}
	e_i^2 & = 0 \, , \\
	e_i \, e_j \, e_i & = e_i \, , \qquad && \text{if} \ |i-j|=1 \, , \\ 
	\bigl[ e_i, e_j\bigr] & = 0 \, ,\qquad && \text{if} \ |i-j|>1 \, .
	\end{aligned}
\end{equation}
This is the usual Temperley--Lieb algebra with `loop fugacity' $\beta = 0$. 
Following \cite{Gainutdinov:2011ab}, we realise this abstract algebra on a fermionic Fock space using \emph{non-unitary} fermionic operators with the following anticommutation relations 
\begin{equation} \label{fermtrans}
	\{f_i,f_j^+\}=(-1)^i\,\delta_{ij} \, , \quad 
	\{f_i,f_j\} = \{f^+_i,f^+_j\} = 0 \, .
\end{equation}
They provide a basis of Fock vectors 
$f_{i_1}^+ \cdots f_{i_M}^+ |\varnothing\rangle$ for the Hilbert space of states.
The free-fermion Temperley--Lieb algebra \eqref{TLrel_freefermion} can then be represented as quadratic expressions in two-site fermionic operators
\begin{equation} \label{TLdef}
	e_k \equiv g_k^+ g_k \, , \qquad 
	g_k \equiv f_k + f_{k+1} \, , \quad g_k^+ = f_k^+ + f_{k+1}^+ \, , 
	\qquad 1\leqslant k < N \, .
\end{equation}

As in \cite{bm2024}, the building blocks of our conserved charges are nested commutators of \emph{adjacent} Temperley--Lieb generators and their (disjoint) anticommutators 
\begin{equation} \label{eq:def nested_TL}
	\begin{gathered}
		e_{[k,l]} \equiv [e_k,[e_{k+1},\ldots[e_{l-1},e_l]\ldots]] = [[\ldots [e_k,e_{k+1}],\ldots e_{l-1}],e_l] \, , \quad  k < l \, , \\
		\{ e_{[i,j]},e_{[k,l]} \} = e_{[i,j]}\,e_{[k,l]} + e_{[k,l]}\,e_{[i,j]} \,, \qquad  i\leqslant j<k\leqslant l \, .
	\end{gathered}
\end{equation}
under the convention $e_{[k,k]} \equiv e_k$.
It can be shown \cite{bm2024} that the nested commutators \eqref{eq:def nested_TL} can be expressed as
\begin{equation}
    e_{[k,l]} = s_{kl} \, \bigl(g_{l}^+ \, g_k + (-1)^{l-k} g_k^+ \, g_{l} \bigr) \, , \qquad k < l \, ,
\end{equation}
where $s_{kl}\equiv(-1)^{(k-l)(k+l-1)/2}$. In particular, all these quantities are quadratic in fermions $f_k$. It follows that the anticommutators of nested commutators $\{e_{[k,l]}, e_{[m,n]}\}$ that we will consider in what follows are at most quartic in fermions.

\runinhead{Definition of the hamiltonians} 

The conserved charges at $q=i$ can be obtained, similarly 
to the odd case \cite{bm2024}, in terms of the numbers $t_k\equiv \tan  \frac{\pi\,k}{N}$.
However, in this case the interaction potentials contain singularities (poles) from the antipodal points with $k=i-j=N/2\equiv L$.
To regularise these poles we introduce a parameter~$\alpha$ and let 
\begin{equation} \label{eq:tangents}
	t_k\equiv \tan  \frac{\pi\,k}{N} \,, \quad k\neq \frac{N}{2} \,, \qquad 
	t_{N/2} =t_{N/2}(\alpha) \equiv \tan  \frac{\pi\,(N/2 +\alpha)}{N} = \cot\frac{\pi\,\alpha}{N} \, .
\end{equation}
The shift by $\alpha$ regularises the simple pole of $t_{N/2}$ at $\alpha= 0$.
From the products
\begin{equation} \label{def t_kl SM}
	t_{k, l} \equiv  \prod_{i=k}^{l-1} t_{i} \, , \quad k<l \, , \qquad t_{k,k} \equiv 1 \, ,
\end{equation}
define
\begin{equation}
	h^{\LL}_{i j}(\alpha) \equiv \! \sum_{n=j+1}^N \!\! t_{n-j,n-i} \, \bigl(1-(-1)^{i} \, t_{n-i,n}^{2}\bigr) \, ,\qquad h^{\RR}_{i j}(\alpha) \equiv(-1)^{i-j} \, h^{\LL}_{N-j, N-i}(\alpha) \, ,
\end{equation}
as well as
\begin{equation} \label{eq:hLRijkl}
	\begin{aligned} 
	h^\LL_{ij;kl} (\alpha) & = (-1)^{k-j} \!\! \sum_{n=l+1}^{N} \!\! t_{n-l,n-j} \; t_{n-k,n-i} \, \bigl(1-(-1)^i \, t^2_{n-i,n} \bigr) \, , \\
	h^\RR_{ij;kl} (\alpha) & = (-1)^{l-j+k-i} \, h^\LL_{N-l,N-k;N-j,N-i} (\alpha) \, .
	\end{aligned}
\end{equation}
These coefficients have at most double poles in $\alpha$ that we remove by taking their residues, 
\begin{equation} \label{eq:hreg}
	\begin{aligned} 
	h_{ij}^+ & \equiv \frac{\pi^2}{2 \, N^2} 
	\lim_{\alpha\to 0} \; \alpha^2 \Big( h^\LL_{ij} (\alpha)\,+\,h^\RR_{ij} (\alpha)\Big) \, , \\
	h_{ij;kl}^- & \equiv \frac{\pi^2}{2 \, N^2} \lim_{\alpha\to 0} 
	\; \alpha^2 \Big( h^\LL_{ij;kl} (\alpha)\,-\,h^\RR_{ij;kl} (\alpha)\Big) \,.
	\end{aligned}
\end{equation}
With these notations, two of the conserved Hamiltonians are given by

\begin{equation} \label{eq:Hams_even}
\begin{aligned} 
	\HH^+ = &
	\sum_{1 \leqslant i \leqslant j < N} \!\!\!\!\! 
	h_{ij}^+ \; e_{[i,j]} \, , \\
	\HH^- = &\! \sum_{1 \leqslant i\leqslant j<k\leqslant l< N} \! 
	h_{ij;kl}^- \, \{ e_{[i,j]},e_{[k,l]} \} \, .
	\end{aligned}
\end{equation}
We can prove that $\HH^+$ and $\HH^-$ are non-vanishing and satisfy
\begin{equation} \label{Hcomm}
	\bigl[ \HH^+ , \HH^- \bigr] = 0 \, .
\end{equation}
This follows from the commutativity of the `chiral' hamiltonians $H^{\LL,\RR}$ of the parent model \cite{Lamers:2020ato} by arguments that are similar to \cite{bm2024}: in short, up to overall factors, $\HH^-$ is the limit of $H^\LL - H^\RR$ as $q \to i$, while $\HH^+$ is the residue of $H^\LL + H^\RR$ at $q = i$.

Like in \cite{bm2024}, we have a quadratic and a quartic hamiltonian, both expressed via nested commutators \eqref{eq:def nested_TL} with coefficients that are products of tangents. Unlike for odd $N$, however, the hamiltonians \eqref{eq:Hams_even} are not diagonalisable, and they are nilpotent: all their eigenvalues vanish. This feature, which is not obvious from the above expressions, will be made more explicit in the examples below.

\runinhead{Other conserved charges} 

In addition to the two Hamiltonians, we get two further commuting quantities by taking the residue of the quasi-translation operator $\G(q)$ and its inverse 
\cite{Lamers:2020ato} as $\G^\pm \equiv \lim_{q \to i} (q-i) \, \G(q)^{\pm 1}$. Explicitly, we have
\begin{equation} \label{eq:q_trans}
    \begin{aligned} 
    	\G^+ & = (1+t_{N-1} \,e_{N-1}) \cdots (1+t_{L+1} \,e_{L+1}) \, e_L\,(1+t_{L-1} \,e_{L-1})\cdots(1+t_{1} \,e_{1})\,, \\
    	\G^- &  = \qquad \ (1-t_{1} \,e_{1}) \cdots (1-t_{L-1} \,e_{L-1})\, e_L\, (1-t_{L+1} \,e_{L+1}) \cdots(1-t_{N-1} \,e_{N-1})\,,
    \end{aligned}
\end{equation}
where we abbreviate $L \equiv N/2$. It follows from the definition of $\G^\pm$ as a limit and the properties of $\G(q)$ that
\begin{equation} \label{eq:qq_trans}
	\G^+ \, \G^- = \G^- \, \G^+ = 0 \, .
\end{equation}
We can show that these charges are conserved
\begin{equation} \label{GHop}
	\bigl[ \G^\pm , \HH^\pm \bigr] = \bigl[ \G^\pm , \HH^\mp \bigr] 	= 0 \, .
\end{equation}

\runinhead{Spectrum}  
Note that $\G^+$ and $\G^-$ are here unsuited for the role of the `quasi-translation' that $\lim_{q \to i} \G(q)$
assumes for odd $N$ \cite{bm2024}, since neither of them is 
invertible. Accordingly, the basis of Fourier transformed fermions $\tilde{\Psi}_i$ in \cite{bm2024} cannot be defined in the same way here. 
All the conserved charges that we have defined here are nilpotent operators, and their spectrum will be given in terms of Jordan blocks. Numerical evidence indicated that at each length there single Jordan block of size $L+1\equiv N/2+1$, and collections of blocks of smaller size.

\runinhead{Symmetries} 

As in the case of the odd-length chain \cite{bm2024}, we can define a parity transformation by $\PP(e_i)=e_{N-i}$. Under this transformation we have 
\begin{equation} \label{parity}
	\PP(\HH^\pm) = \pm\HH^\pm\, , \quad \PP(\G^\pm) = \G^\mp \, .
\end{equation}
Note that as opposed to the odd-length case, the quadratic hamiltonian is here the one that is parity invariant while the quartic hamiltonian is anti-invariant.

It follows from the presentation of the model in terms of the Temperley--Lieb algebra that the model possesses a global $\mathfrak{gl}(1|1)$-symmetry given by
\begin{equation} \label{global_gl11}
	\begin{gathered}
	\F_1 = \sum_{i=1}^N f_i \, , \quad \F_1^+ = \sum_{i=1}^N f^+_i \,, \quad
	\N = \sum_{i=1}^N \, (-1)^i \, f^+_i \mspace{1mu} f_i \, , \quad \EE =
	\sum_{i=1}^N \, (-1)^i = 0 \, .
	\end{gathered}
\end{equation}
The vanishing of the central charge $E$ is specific to even $N$ and signals the occurrence of indecomposable features in the representations, hence the Jordan blocks \cite{Gainutdinov:2011ab}. 
The $\frak{gl}(1|1)$ symmetry can only account for Jordan blocks of size two, and is only part of the symmetry, as can be concluded from the $U_q(\widehat{\mathfrak{sl}}_2)$ action for the parent model with generic $q$. The limit to $q=i$ of this algebraic structure will be considered in more detail in a future paper.

In the next section we will give examples of the action of the conserved charges and the symmetry algebra for systems of small sizes.

\section{Small-size examples}
\label{sec:3}

In this section we illustrate the limiting procedure $q\to i$ on the smallest even size $N=2$ and we give the conserved quantities and the structure of the Hilbert space for the smallest few sizes.

\runinhead{Origin from \textit{q}HS for \textit{N} = 2} 

It is instructive to consider $N=2$ in some detail. Here, the hamiltonians $\HH^{\textsc{l}}(q)$ and $\HH^{\textsc{r}}(q)$ defined in \cite{Lamers:2020ato} for general $q$ are equal, up to normalisation, to the Temperley--Lieb generator
\begin{equation}
    e(q) = \begin{pmatrix} 
	\, 0 & \color{gray!80}{0} & \! \color{gray!80}{0} & \color{gray!80}{0} \, \\
	\, \color{gray!80}{0} & q^{-1} & -1 & \color{gray!80}{0} \, \\
	\, \color{gray!80}{0} & -1 & q & \color{gray!80}{0} \, \\
	\, \color{gray!80}{0} & \color{gray!80}{0} & \color{gray!80}{0} & 0 \, \\
	\end{pmatrix} \, ,
\end{equation}
here expressed in the computational basis $\ket{\uparrow\uparrow},\ket{\uparrow\downarrow},\ket{\downarrow\uparrow},\ket{\downarrow\downarrow}$ of $\mathbb{C}^2 \otimes \mathbb{C}^2$. 
More precisely
\begin{equation}
	\HH^+(q) = \HH^{\textsc{l}}(q) + \HH^{\textsc{r}}(q) = \frac{e(q)}{q+q^{-1}} \,, \qquad 
	\HH^-(q) = \HH^{\textsc{l}}(q) - \HH^{\textsc{r}}(q) =0 \,.
\end{equation}
$\HH^+(q)$ has eigenvalues $\{0^3,1\}$ with eigenbasis basis formed by the $q$-symmetric states $ \ket{\uparrow\uparrow}$, $\ket{\uparrow\downarrow} + \, q^{-1} \ket{\downarrow\uparrow}$, $\ket{\downarrow\downarrow}$ and the $q$-antisymmetric state $\ket{\uparrow\downarrow}- q \, \ket{\downarrow\uparrow}$.
When $q\to i$ the matrix elements diverge while the energies remain finite, and the $q$-symmetric and $q$-antisymmetric one-magnon states become collinear. In order to remove this pathology one can rescale the hamiltonian as
\begin{equation}
\label{n=2}
	\HH^+=\lim_{q\to i} \; (q+q^{-1})\,\HH^+(q) = \lim_{q\to i} \,e(q)\,,
\end{equation}
which sends all the eigenvalues to $0$.

\runinhead{Results for low \textit{N}}

After rescaling and sending $q$ to $i$ we find it more convenient to use the Fock basis associated to the fermions defined in \eqref{fermtrans}. Write $M$ for the number of fermions.

\runinhead{{\textbf{\itshape N} = 2.}} 
In the fermionic description the hamiltonian \eqref{n=2} for the smallest length is
\begin{equation}
	\HH^+ = \G^\pm = e_1=(f_1^++f_2^+)(f_1+f_2)\,, \qquad \HH^- =0\,.
\end{equation}
The vectors in the representation are given by 
\begin{equation}
	\ket{\varnothing}\,, \quad (f_1^++f_2^+) \, \ket{\varnothing}\,, \quad (f_1^+-f_2^+) \, \ket{\varnothing} \,, \quad f_1^+ f_2^+ \, \ket{\varnothing} \,.
\end{equation}
which form a four-dimensional indecomposable representation of $\mathfrak{gl}(1|1)$. 
$\HH^+$ has a $2\times 2$ Jordan block in the $M=1$ sector. 

\runinhead{{\textbf{\itshape N} = 4.}} The conserved quantities are given by
\begin{equation} \label{N4 ex}
\begin{aligned} 
	\G^\pm & = (1-t_1 \, e_{2\pm 1}) \, e_2 \, (1+t_1 \, e_{2\mp 1 })\,, \\
	\HH^+ & =[e_1+e_3,e_2]+e_1+e_3-2\,e_2\, , \\
	\HH^- & = -2\{e_1-e_3,e_2\} - \{e_1,[e_2,e_3]\} - \{[e_1,e_2],e_3\} \\
	& = -2 \bigl( \{e_1-e_3,e_2\} + e_1 \, e_2 \, e_3 - e_3 \, e_2 \, e_1 \bigr)\, .
\end{aligned} 
\end{equation}
Both $\HH^+$ and $\HH^-$ have the following Jordan-block structure: a $1\times 1$ block at $M=0$ (and $M=4$), two $2\times 2$ blocks at $M=1$ (and at $M=3$), and a $3\times 3$ block plus three $1\times1$ blocks at the equator $M=2$.
\medskip

In general, $\G^\pm$ have Jordan blocks up to size $M+1$ at each $M\leqslant N/2$. For $N\geqslant 6$ the Jordan-block structures of $\HH^\pm$ do not necessarily coincide.

\begin{acknowledgement}
We thank A.~Toufik for collaboration at an early stage of this work. JL was funded by ERC-2021-CoG\,--\,BrokenSymmetries 101044226. DS thanks the organisers of the program \textit{Mathematics and Physics of Integrability} (MPI2024) held in July 2024 at \textsc{Matrix} in Creswick \textsc{Vic}, Australia,  for hospitality, and SMRI for financial support.
\end{acknowledgement}

\bibliographystyle{spmpsci.bst}
\bibliography{Mybib} 

\end{document}